\documentclass{nature}[a4paper]
\usepackage{subcaption} 
\usepackage{caption} 
\usepackage{float}
\usepackage{array,multirow}
\usepackage{booktabs}
\usepackage{nameref}
\usepackage{xcolor,colortbl}
\usepackage{url}
\usepackage{hyperref}

\definecolor{Gray}{gray}{0.85}

\renewcommand{\arraystretch}{1.5}

\usepackage{authblk}
\usepackage{amsmath}
\usepackage{graphicx}
\usepackage{xcolor}
\makeatletter
\let\saved@includegraphics\includegraphics
\AtBeginDocument{\let\includegraphics\saved@includegraphics}
\renewenvironment*{figure}{\@float{figure}}{\end@float}
\makeatother
\usepackage[labelfont=bf]{caption}
\usepackage{cite}
\newcommand{\upcite}[1]{\textsuperscript{\scalebox{0.7}{\cite{#1}}}}

\title{FerroX : A GPU-accelerated, 3D Phase-Field Simulation Framework for Modeling Ferroelectric Devices \footnotetext{$^\ast$  E-mail: prabhatkumar@lbl.gov, jackie\_zhiyao@lbl.gov}}

\author[1,$\ast$]{Prabhat Kumar}
\author[1]{Andrew Nonaka}
\author[1]{Revathi Jambunathan}
\author[2]{Girish Pahwa}
\author[3,4]{Sayeef Salahuddin}
\author[1, $\ast$]{Zhi (Jackie) Yao}

\affil[1]{Applied Mathematics and Computational Research Division, Lawrence Berkeley National Laboratory, CA, 94720, USA}
\affil[2]{Berkeley Device Modeling Center, University of California, Berkeley, CA 94720, USA}
\affil[3]{Materials Sciences Division, Lawrence Berkeley National Laboratory, CA 94720, USA}
\affil[4]{Department of Electrical Engineering and Computer Sciences, University of California, Berkeley, CA, 94720, USA}

\graphicspath{{figures/}}
\date{}                   
\begin{document}
	
	\maketitle
	\vspace{10pt}

\begin{abstract}
\textbf{We present a massively parallel, 3D phase-field simulation framework for modeling ferroelectric materials based scalable logic devices. 
This code package, FerroX, self-consistently solves the time-dependent Ginzburg Landau (TDGL) equation for ferroelectric polarization, Poisson's equation for electric potential, and charge equation for carrier densities in semiconductor regions. The algorithm is implemented using the AMReX software framework \cite{zhang2021amrex}, which provides effective scalability on manycore and GPU-based supercomputing architectures. We demonstrate the performance of the algorithm with excellent scaling results on NERSC multicore and GPU systems, with a significant (15x) speedup on the GPU using a node-by-node comparison. We further demonstrate the applicability of the code in simulations of ferroelectric domain-wall induced negative capacitance (NC) effect in Metal-Ferroelectric-Insulator-Metal (MFIM) and Metal-Ferroelectric-Insulator-Semiconductor-Metal (MFISM) devices. The charge (Q) v.s. voltage (V) responses for these 3D structures clearly indicates stabilized negative capacitance with multidomain formation, which is corroborated by amplification of the voltage at the interface between the ferroelectric and dielectric layers.}
\end{abstract}




\maketitle

Due to their switchable polarization in response to applied electric fields, ferroelectric materials have enabled a wide portfolio of innovative microelectronics devices, such as ferroelectric capacitors (FeCaps) \upcite{Miller1990, Zhang2020}, ferroelectric tunnel junctions (FTJs) \cite{Y.2006} and ferroelectric field effect transistors (FeFET) \cite{Khan2020, Aziz2018, Mikolajick2021, Mikolajick2020}. 
The remnant polarization in the ferroelectric material at zero applied electric field allows for nonvolatile retention in these devices \cite{Mikolajick2021}.
The unique physics of FeFETs has been instrumental in the design of other new technologies including nonvolatile memories \cite{Eaton1988, Kohlstedt2005}, logic-in-memory (LiM) architectures \cite{Breyer2019, Yin2019}, oscillators and spiking neurons \cite{Wang2018, Khan2020} and negative capacitance field effect transistors (NCFETs) \cite{salahuddin2008use, khan2015negative, islam2011experimental, Yadav2019, cheema2022ultrathin, Lukyanchuk2022}. 
NCFETs, in particular, are designed to overcome the fundamental energy consumption limit (the `Boltzmann's Tyranny' \cite{D.2001}) associated with individual semiconductor components \cite{salahuddin2008use, Zubko2016, Wong2018}, allowing for the design of ultra low-power logic technologies.
The negative capacitance manifestations have been experimentally observed both macroscopically \cite{Appleby2014, Jo2016, McGuire2017, Hoffmann2022} and locally \cite{Yadav2019}.
With more groundbreaking discoveries on the way, we expect the field to culminate in the demonstration of highly-integrated, CMOS-compatible, ultra-low-power, sub-10-nanometer, longer-retention and higher-endurance ferroelectric ecosystems. 

Modeling and simulation is playing an increasing role in providing in-depth insights into the underlying physics, as well as in paving the road to facilitate researchers with reliable design tools for new microelectronic devices \cite{chen2008phase}.
One of the key challenges in the modeling of devices such as FeFETs and NCFETs is the intrinsic multiphysics nature of the multimaterial stacks.
Typical ferroelectric devices involve at least three coupled physical mechanisms: ferroelectric polarization switching, semiconductor electron transport, and classical electrostatics, each of which includes rich underlying physics.
Therefore, accurate numerical coupling schemes are essential for studying and designing ferroelectric devices. 
Current state-of-the-art modeling works have demonstrated some coupling of the entire system of interest.
The prevalent phase field modeling method has been applied to study polarization dynamics in ferroelectric materials, usually stacked with dielectric layers as in FeFETs \cite{ashraf2012phase, Zubko2016, li2001phase,li2005ferroelectric,li2006temperature,li2008influence,lee2006ferroelastic,chen2008phase,chen1998applications, Lukyanchuk2022}, but without consideration of the charge distributions in semiconductor substrates. 
Others have also considered strain effects \cite{chen2008phase, ashraf2012phase, li2001phase, li2005ferroelectric, li2006temperature, li2008influence} and temperature effects \cite{chen2008phase,li2001phase, li2005ferroelectric, li2006temperature, li2008influence, Zubko2016}. 
Recent works have included the semiconductor charges, but in a two-dimensional (2D) geometrical representation \cite{saha2020multi, pahwa2019numerical}. 
However, full three-dimensional (3D) considerations of the structures are required to accurately capture the complex heterogeneous structure operations, especially in miniature designs such as ferroelectric finFET \cite{Krivokapic2017, Seo2018, Choi2001,Yu2002}.
This 3D aspect sets additional requirements on the mesh resolution, making the simulation efficiency difficult to realize with complicated and coupled physical processes. 
Therefore, an accurate and efficient 3D simulation tool to model ferroelectric-dielectric-semiconductor heterostructures, that is portable from laptops to manycore/GPU exascale systems, is in urgent demand.

In this article, we present a massively parallel, 3D phase-field simulation framework, named FerroX, for modeling and design of ferroelectric-based microelectronic devices.
The overall strategy couples the time-dependent Ginzburg Landau (TDGL) equation for ferroelectric polarization, Poisson's equation for electric potential, and charge equation for carrier densities in semiconductor regions. 
We discretize the coupled system of partial differential equations using a finite difference approach, with an overall scheme that is second-order accurate in both space and time.
Boundary condition options for various surface effects (free polarization, zero polarization and finite extrapolation length $\lambda$) on the polarization at ferroelectric-dielectric and metal-ferroelectric interfaces have been implemented. 
In order to achieve a massively parallel manycore/GPU implementation of our structured grid simulations, we leverage the DOE Exascale Computing Project (ECP) code framework, AMReX, developed by Zhang et al. \cite{zhang2021amrex}. 
We demonstrate the applicability of our code with simulations of ferroelectric domain-wall induced negative capacitance (NC) effects in Metal-Ferroelectric-Insulator-Metal (MFIM) and Metal-Ferroelectric-Insulator-Semiconductor-Metal (MFISM) structures. 
Considering hafnium–zirconium oxide (Hf$_{0.5}$Zr$_{0.5}$O$_2$, or HZO) as the ferroelectric material, we study the dynamics of domain walls and polarization switching in MFIM and MFISM stacks using our validated 3D code. 
The charge (Q) v.s. applied voltage (V) responses for these structures clearly indicates stabilized negative capacitance. 
We also demonstrate the performance of the algorithm with excellent scaling results on NERSC multicore and GPU systems, with a significant (15x) speedup on the GPU using a node-by-node comparison \upcite{nerscwebsite}.

\section*{Results}\label{result}
\subsection{Overview of phase-field simulation framework}
\begin{figure}
    \centering
    \includegraphics[width=\linewidth]{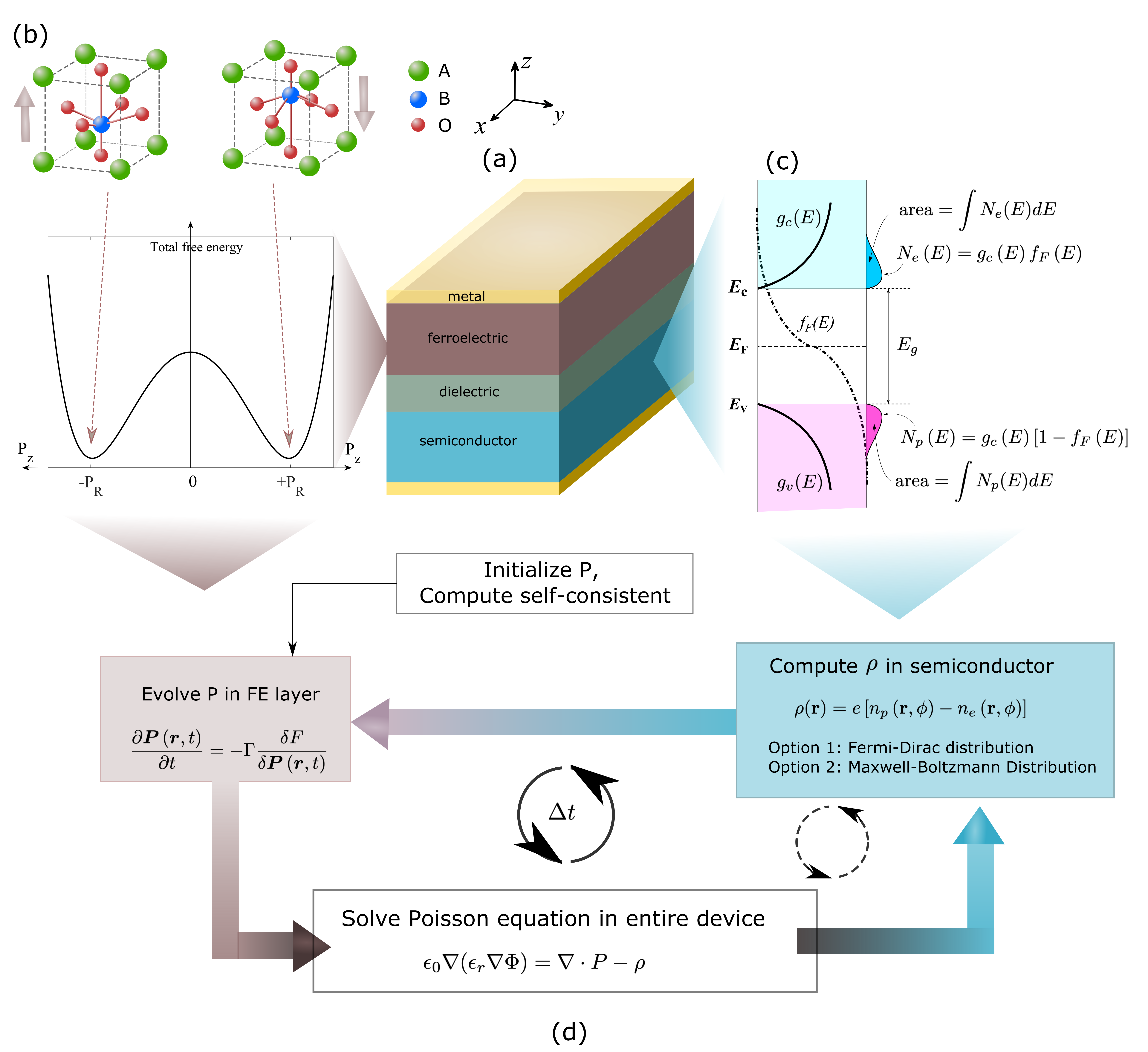}
    \caption{Overview of FerroX, a GPU-enabled phase-field simulation framework. (a) Schematic of an MFISM stack. Applied voltage across the stack is controlled by specifying electric potentials on the top and bottom metal contacts shown in yellow. Removing the semiconductor layer reduces the structure into an MFIM stack. (b) Uniaxial atomic displacement along the thickness (z-direction) of the ferroelectric film and corresponding double-well energy landscape. (c) Carrier concentration in semiconductor region is described using Fermi-Dirac distribution function as function of potential distribution in the device. (d) A typical time-step of FerroX. 
    We iteratively solve Poisson's equation and compute $\rho$  until self-consistency is achieved. 
    See Section~ ``\nameref{sec:model}'' in Methods.}
    \label{fig:overview}
\end{figure}

Fig.  \ref{fig:overview} provides an overview of the physical model and numerical approach in FerroX.
Here we consider an MFISM stack comprising of ferroelectric (FE), dielectric (DE), and semiconductor (SC) thin films as shown in Fig.~\ref{fig:overview}(a),  which is the essential functioning block in FeFETs and NCFETs. 
The applied voltage across the device is controlled by specifying the electric potentials on top and bottom metal contacts shown in yellow. The dynamics of polarization in ferroelectrics is described by the time-dependent Ginzburg–Landau (TDGL) equation~\cite{chen2008phase,saha2019phase,saha2020multi,park2019modeling}:
\begin{equation}
    \frac{\partial\mathbf{P}(\mathbf{r},t)}{\partial{t}} = -\Gamma\frac{\delta F}{\delta\mathbf{P}(\mathbf{r},t)}
\label{eq:TDGL1}
\end{equation}
where $\mathbf{P} = (P_x, P_y, P_z)$ is the electric polarization vector, $\mathbf{r} = (x,y,z)$ is the spatial vector, $\Gamma$ is the kinetic or viscosity coefficient, and $\frac{\delta F}{\delta\mathbf{P}(\mathbf{r},t)}$ represents the driving force for the evolution of system. 
$F$ is the total free energy of system as a function of $\mathbf{P}(\mathbf{r},t)$ and takes into account the contributions due to the bulk Landau free energy, the domain wall energy or gradient energy, and the electric energy of the applied electric field. 
Exact forms of these energy densities are elaborated in the  ``\nameref{subsec:TDGL}'' section in the Methods.
In ferroelectric materials, spontaneous displacement of atoms leads to a non-zero spontaneous polarization. 
In most FeFETs, atomic displacement is dominantly uniaxial along the out-of-plane direction (z-direction in Fig.~\ref{fig:overview}(b)), therefore, in this article, we assume the polarization vector to present only a out-of-plane component, i.e. $P_x = P_y = 0$ and $P_z = P$. 

The distribution of electric potential in the system is obtained by solving Poisson's equation in the following form:
\begin{equation}
    \nabla\cdot\epsilon\nabla\Phi = \nabla\cdot \mathbf{P} - \rho
\end{equation}
where $\epsilon$ is a spatially-varying permittivity and $\rho$ is the total free charge density in the semiconductor region and can be set to zero for MFIM devices since a semiconductor layer is not present.  On the other hand, for devices with a semiconductor layer, charge density in the semiconductor region depends on the local distribution of electric potential and is computed using the following equation~\cite{neamen2003semiconductor}:
\begin{equation}
    \rho(\mathbf{r}) = e\left[n_p - n_e + N_d^+ - N_a^-\right]
\label{eq:rho}
\end{equation}
where $e$ is the elementary charge and $n_p(\mathbf{r})$, $n_e(\mathbf{r})$, $N_d^+(\mathbf{r})$, and $N_a^-(\mathbf{r})$ are densities of holes, electrons, ionized donors, and acceptors at spatial location $\mathbf{r}$. $N_d^+(\mathbf{r})$ and $N_a^-(\mathbf{r})$ are negligible if we consider undoped silicon as the semiconductor layer. Electron and hole densities at equilibrium can be estimated using Fermi-Dirac statistics as shown in the schematic in Fig.~\ref{fig:overview}(c).

A typical time step of FerroX is shown in Fig.~\ref{fig:overview}(d). We initialize polarization $\mathbf{P}$ using a uniformly distributed random numbers and compute the corresponding distributions of potential $(\Phi)$ and charge density $(\rho)$ iteratively until self-consistency is reached. At each time step we solve the TDGL equation to update $\mathbf{P}$ and use the updated $\mathbf{P}$ to compute $\Phi$ and $\rho$. Since the right-hand-side of the Poisson's equation is also a function of $\Phi$, we use fixed-point iteration, where we iteratively lag the effects of $\Phi$ when computing the right-hand-side.  
The iterations are stopped when the average magnitude of change over all cells is less than a user-defined tolerance. For the applications described in this paper, the tolerance is $1\times 10^{-5}$. The Poisson solver used in our framework provides second order accuracy in space. 
We have designed the temporal integrators for the TDGL equation to produce either first or second-order convergence in time with an option to choose either at run time. 
We should note that our spatial discretization is second-order within each material, however, discontinuities in our model near interface boundaries will result in a reduction in the order of convergence in space (see the ``\nameref{sec:numericalvalidation}'' section in the Methods). 

\subsection{Scaling performance on GPUs and CPUs}
\begin{figure}[htb]
    \centering
    \includegraphics[width=0.6\linewidth]{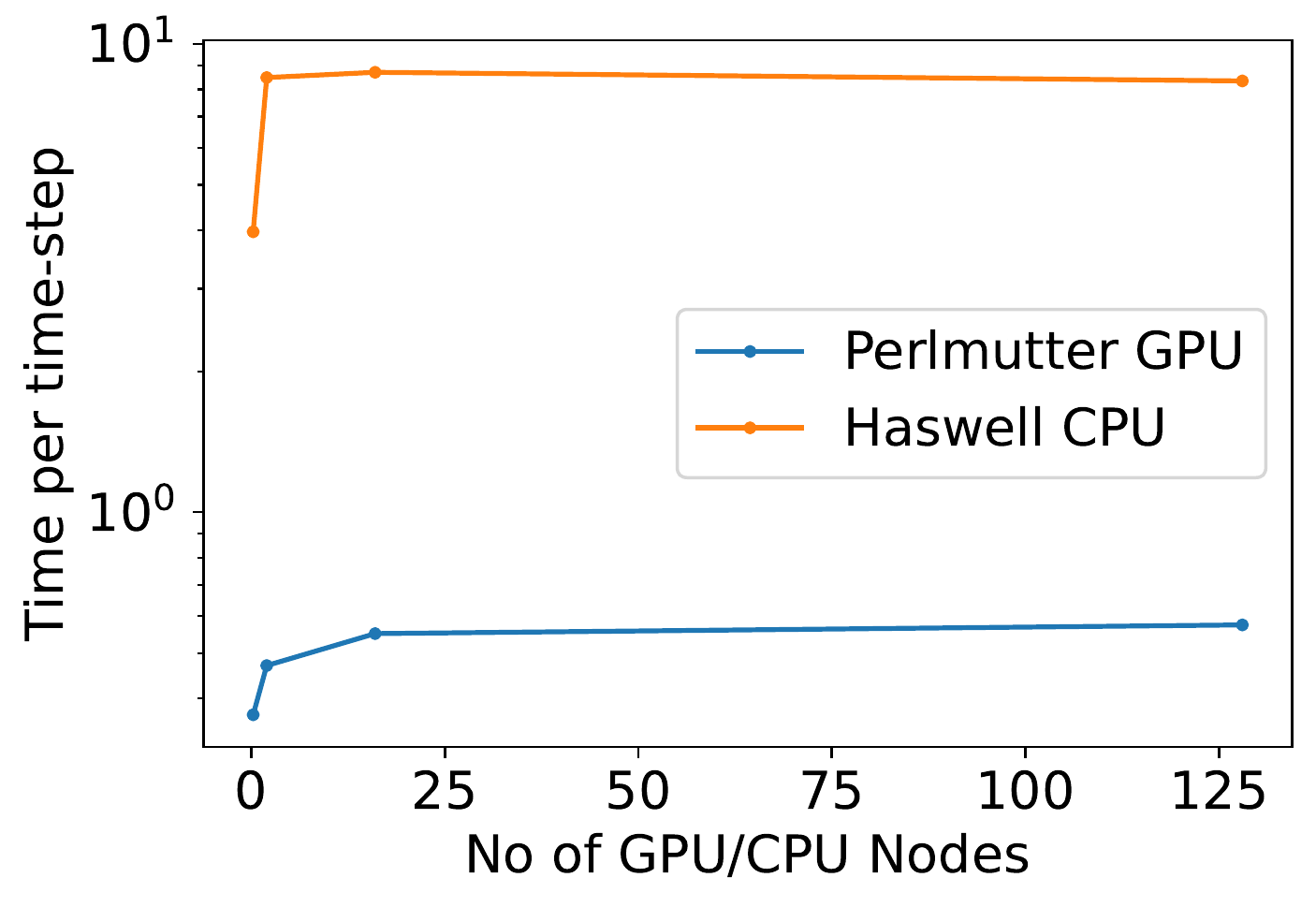}
    \caption{Weak scaling for simulations on the NERSC Cori Haswell (MPI-only) and Perlmutter GPU (MPI+CUDA) systems, respectively. Note that one Haswell node contains 32 physical cores, and one Perlmutter node contains 4 GPUs. The weak scaling is nearly perfect for the Perlmutter simulations past the 1-node threshold up to 512 GPUs (128 nodes), as well as for the Haswell partition simulations past the 1-node threshold up to 4096 CPUs (128 nodes). Also, when using 2 nodes or more, the GPU simulations run 15x faster than the pure MPI simulations on a node-by-node comparison. Details of numerical accuracy and performance tests can be found in the ``\nameref{sec:scaling}'' section in the Methods.}
    \label{fig:scaling}
\end{figure}

FerroX is massively parallel and is performance-portable from laptops to manycore/GPU exascale systems, with a significant speedup on the NERSC GPUs compared to the NERSC CPUs on a node-by-node comparison \upcite{nerscwebsite}. 
We performed weak scaling studies, holding the number of grid points per core/GPU constant as we increase the problem size.
The weak scaling results in Fig.~\ref{fig:scaling} indicate the average simulation time per time step as a function of the number of nodes, on both NERSC Perlmutter GPUs and NERSC Cori Haswell CPUs. 
We performed tests on each system up to 128 nodes, which corresponds to 4096 CPUs on Haswell and 512 GPUs on Perlmutter. 
For the 1/4th node baseline runs, the GPU simulations performed 11x faster than the pure MPI simulations on the Haswell partition. 
We also observe that for the 2-node through 128-node runs, the GPU simulations performed approximately 15x faster than the pure MPI simulations on the Haswell partition. 
Both the Haswell partition and Perlmutter GPU simulations achieve nearly perfect scaling up to 128 nodes beyond the 1-node threshold. 
The excellent parallel performance enables FerroX to adequately address realistic ferroelectric design challenges with full physical and geometrical details, which were previously disregarded by available commercial or research grade software, due to the overwhelming computational complexity.

\begin{table}
\centering
\begin{tabular}{l  l  r}\toprule
Parameters & Values & Units \\
\\[-1.8em]
\hline
\rowcolor{Gray}  $\alpha$ & $-2.5\times 10^9$ & $~\rm{Vm/C}$ \\
\\[-1.8em]
    $\beta$ & $6.0\times 10^{10}$ & $~\rm{Vm^5/C^3}$\\
\\[-2.0em]
\rowcolor{Gray}    $\gamma$ & $1.5\times 10^{11}$ & $~\rm{Vm^9/C^5}$\\
    $g_{11} = g_{44}$ & $1.0\times 10^{-9}$ & $~\rm{Vm^3/C}$\\
\\[-2.0em]
\rowcolor{Gray}    $\epsilon_\mathrm{\rm fe}$ & $24.0$ & 1\\
\\[-2.0em]    
    $\epsilon_\mathrm{DE}$ &  $10 ({\rm{Al}_2\rm{O}_3)}$,  $3.9 (\rm{SiO}_2)$ & 1\\
\\[-2.0em]
\rowcolor{Gray}    $\epsilon_\mathrm{SC}$ & $11.7$ & 1\\
\\[-2.0em]
$m_{e}^*$ & $1.08\times m_e$ & $\mathrm{kg}$ \\
\\[-1.8em]
\rowcolor{Gray}    $m_{p}^*$ & $0.81\times m_e$ & $\mathrm{kg}$ \\
\\[-2.0em]
    $\Gamma$ & $100.0$ & 1 \\
\\[-2.0em]
\rowcolor{Gray}    $\Delta x = \Delta y = \Delta z$ & $0.5 \times 10^{-9}$ & $~\rm{m}$ \\
\\[-2.0em]
$\lambda$ & $3.0\times 10^{-9}$ & $\mathrm{m}$ \\
\\[-1.8em]
\rowcolor{Gray}    $\Delta t$ & $4.0\times 10^{-13}$ & $~\rm{s}$ \\
[-0.3em]\bottomrule
\end{tabular}
\caption{Physical and numerical parameters used in the simulations}
\label{tab:sim_param}
\end{table}

\subsection{Domain dynamics and NC effect in MFIM stacks}
Using FerroX, we explore the dynamics of domain walls, polarization switching, and the negative capacitance effect in an MFIM stack.
The device consists of a 5-nm-thick HZO and 4-nm-thick Al$_2$O$_3$ as the ferroelectric and dielectric layers, respectively. Lateral dimensions along the $x$ and $y$ axes are 32 nanometers. 
We perform the simulations in both 2D and 3D Cartesian geometry. 
2D simulations are performed with parameters similar to that used by Saha and Gupta~\cite{saha2020multi}, and serve as a validation case for our algorithm. 
To realize a 2D structure, we initialize the polarization $P$ in the ferroelectric film to be uniform along one of the lateral dimensions $(y)$ in a 3D FerroX simulation, which effectively turns off the contribution from gradients in $P$ and $\Phi$ along the $y$ direction,
so the problem reduces to two dimensions in the $x-z$ plane. 
In contrast, 3D cases are initialized with a random distribution of $P$ along all three dimensions, while keeping the same physical sizes and material properties as the 2D cases. 

\begin{figure*}
    \centering
    \includegraphics[width=0.75\linewidth]{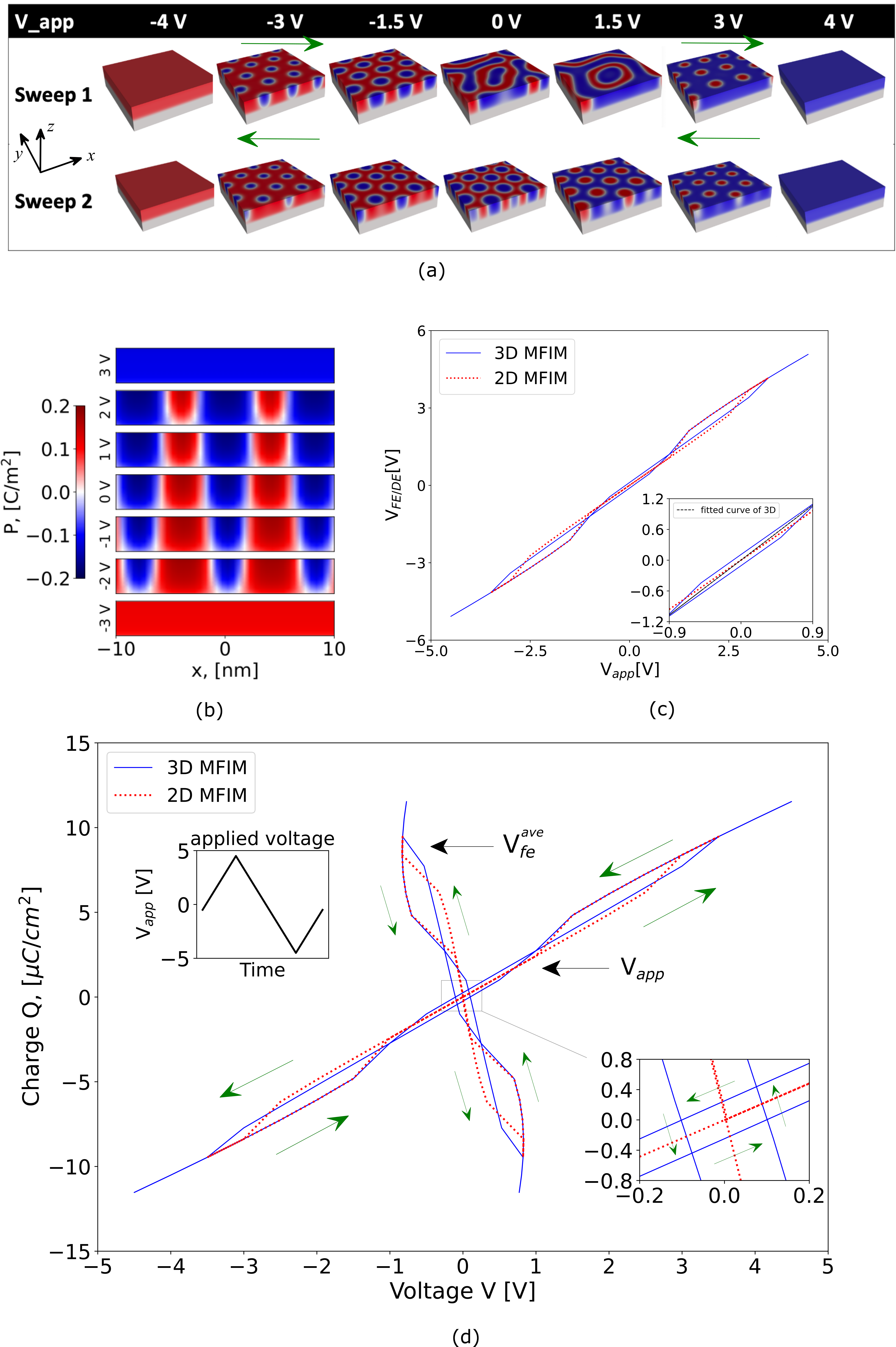}
    \caption{\small Polarization switching dynamics and $Q-V$ characteristics in MFIM stack using 2D and 3D models. 
    5 nm thick HZO and 4 nm thick Al$_2$O$_3$ are used as the ferroelectric and dielectric layers with physical and numerical parameters as shown in Table \ref{tab:sim_param}. 
    (a) and (b) polarization domains in 3D and 2D respectively, 
    (c) $V_{FE-DE}^{\mathrm avg}-V_\mathrm{app}$, where $V_{FE-DE}^{\mathrm avg}$ is the average potential at the FE-DE interface, and 
    (d) $Q-V$ characteristics. For $V_{\rm app} = 0$ V, FE is in multidomain state with approximately equal positive and negative domains in either case (panel labeled 0 V in (a) and (b)), however, with different domain structures in the 2D and 3D scenarios. 
    Note that the dielectric layer in 2D is not shown. 
    As the applied voltage is varied, following the triangular wave profile as shown in the top inset of (d), FE transitions from multidomain to single-domain (either $P\downarrow $ or $P\uparrow $).
    This voltage sweep process is repeated multiple times, with the first sweep (``Sweep 1'' in (a)) to realize the first poling of FE.
    Corresponding $Q-V$ characteristics exhibit double hysteresis. Simulations show that transition from multidomain to single domain occurs at a smaller applied voltage in 2D compared to that in 3D.}
    \label{fig:MFIM}
\end{figure*}

Fig.~\ref{fig:MFIM} shows applied-voltage-induced polarization switching due to the dynamics of domain walls, as well as the negative capacitance effect in an MFIM stack.
First, we investigate multidomain formation in FE by considering an applied voltage $V_{\rm app}$ of 0 V across the stack. 
Once initialized, the polarization $P$ in the ferroelectric layer evolves towards equilibrium with minimized total energy.
During this relaxation process, the interaction between various energy terms (given in equation (\ref{eq:free}), (\ref{eq:grad}) and (\ref{eq:elec}) in the Method section) dictates the formation and motion of the ferroelectric domains. 
In the presence of a DE layer, connected in series with the FE layer, the electric field seen by the FE layer is given by
    $E_{z,\mathrm{\rm dep}} = {-P}\slash{\left[\epsilon_0\left(\epsilon_\mathrm{\rm fe}+\epsilon_\mathrm{DE}\frac{t_\mathrm{\rm fe}}{t_\mathrm{DE}}\right)\right]}$
for zero applied voltage \cite{alam2019positive}. 
Here $\epsilon_\mathrm{FE(DE)}$ and $t_\mathrm{FE(DE)}$ are the relative permittivity and thickness of FE (DE), respectively; $\epsilon_0$ is the vacuum permittivity, and $P$ is the polarization in FE. 
This electric field, $E_{z,\mathrm{\rm dep}}$, is the depolarization field, which opposes the spontaneous polarization $P$ and results in an increase in the depolarization energy term.
This increase is compensated by the formation of $180^\circ$ soft domains of alternate positive and negative $P$ values, as shown in the domain patterns corresponding to $V_{\rm app} = 0$~V in Fig.~\ref{fig:MFIM}(a) and (b).
Note that both these figures show the domains at steady state. 
 
The $Q-V$ characteristics of the device is shown in Fig.~\ref{fig:MFIM}(c) and (d). 
Specifically, Fig.~\ref{fig:MFIM} (c) shows the $Q - V_{\rm fe}^{\rm avg}$ relation and Fig.~\ref{fig:MFIM}(d) shows the $Q-V_{\rm app}$ relation. 
Here, the average charge density $Q$ at the FE-DE interface $z = z_{\rm int}$ is calculated as
$Q(z_{\rm int}) = \frac{1}{L_xL_y}\int_0^{L_x}\int_0^{L_y}\epsilon_0\epsilon_{DE}\times E_{z_{\rm int}}dxdy $
where $L_x$ and $L_y$ are lateral dimensions of the film and $E_{z_{\rm int}}$ is the electric field on the dielectric side of the $x-y$ plane at $z = z_{\rm int}$. 
The average voltage drop across the FE layer, denoted as $V_{\rm fe}^{\rm avg}$, is computed as
$   V_{\rm fe}^{\rm avg} = V_{\rm app} - V_{\rm FE/DE}^{\rm avg} = V_{\rm app} - \frac{1}{L_xL_y}\int_0^{L_x}\int_0^{L_y}\Phi_{\rm int}dxdy, $
 where $\Phi_{\rm int}$ is the potential on the $x-y$ plane at $z = z_{\rm int}$. 
 As the applied voltage increases from $0~\rm{V}$, indicated by the first quarter period of the triangle wave profile in the top inset of Fig.~\ref{fig:MFIM}(d), regions of negative $P$ (denoted as $P\downarrow$) enlarge in size and regions of positive $P$ (denoted as $P\uparrow$) reduce in size due to the domain wall motion.
 This process can be observed in both the 3D and 2D structures, as shown in the 3D domain patterns in Fig.~\ref{fig:MFIM}(a) ( $V_{\rm app}$ of 0 V, 1.5 V, 3 V and 4V),  and the 2D domain patterns in Fig.~\ref{fig:MFIM}(b) ($V_{\rm app}$ of 0 V, 1 V, 2 V and 3V).
 The FE layer eventually goes into a  $P\downarrow$ single-domain with $V_{\rm app} = V_{\rm SD}  \sim 3.5 V$ for the 3D structure and $V_{\rm app} = V_{\rm SD} \sim 3.0 V$ for the 2D structure.
 Once single-domains are formed, we start decreasing the applied voltage and our simulations show that the FE layer returns to a multidomain state when $V_{\rm app}$ gets to $V_{\rm MD} \sim 1~\rm{V}$. 
 The double-well energy landscape induces a hysteresis loop in the  $Q-V_{\rm app}$ characteristic curve \cite{saha2020multi,LandauerParadox}, which is seen in the first quadrant in Fig.~\ref{fig:MFIM}(d).
Similarly, as applied voltage $V_{\rm app}$ is further reduced below $ 0~\rm{V}$, regions of $P\uparrow$ enlarge in size and regions of  $P\downarrow$ reduce in size and the FE layer eventually goes into a $P\uparrow$ single-domain at $V_{\rm app} = -V_{\rm SD}$. 
The applied voltage is then increased, which results in FE returning to the multidomain state, exhibiting the ``second'' hysteresis loop in the third quadrant of the $Q-V_{\rm app}$ relation in Fig.~\ref{fig:MFIM}(d). 
 
The negative slope of the $Q-V_{\rm fe}^{\rm avg}$ curve in Fig.~\ref{fig:MFIM} (d) indicates a negative capacitance $C_{\rm fe}^{\rm avg} = dQ/dV_{\rm fe}^{\rm avg}$ when the FE is in multidomain state. 
 One can deduce from this NC property that the average effective permittivity has a negative $z$ component $(\epsilon_{z,{\rm fe}})$.
To quantify this effect, we define the strength of the NC effect as $\lvert1/(\epsilon_0\epsilon_{z,{\rm fe}})\rvert = \lvert dE_{z,{\rm fe}}^{\rm avg}/dQ\rvert$~\cite{saha2020multi,iniguez2019ferroelectric,Zubko2016}. Based on this definition, it can be inferred from Fig.~\ref{fig:MFIM} (d) that the NC effect is stronger in 3D compared to that in 2D (since the 2D curve is steeper).
The stronger NC effect in 3D is also revealed by the larger differential amplification $dV_{\rm FE/DE}^{\rm avg}/dV_{\rm app}$ shown in Fig.~\ref{fig:MFIM} (c).
Such enhancement of NC effect can be attributed to higher density of domain walls in 3D.
Since NC effect is induced by domain wall motion, thus it is affected by the domain wall energy density, $F_\mathrm{dw} = f_{x,\mathrm{elec}} + f_{y,\mathrm{elec}} + f_{x,\mathrm{grad}} + f_{y,\mathrm{grad}}$ ~\cite{Yadav2019}, where
$f_{x,\mathrm{elec}} = \epsilon_0\epsilon_xE_{x,FE}^2$, $f_{y,\mathrm{elec}} = \epsilon_0\epsilon_yE_{y,FE}^2$, $f_{x,\mathrm{grad}} = \frac{1}{2}g_{44}(d^2P/dx^2)$, and $f_{y,\mathrm{grad}} = \frac{1}{2}g_{44}(d^2P/dy^2)$.
The 2D assumption artificially underestimates $F_\mathrm{dw}$ due to the absence of the $ f_{y,\mathrm{elec}}$ and  $f_{y,\mathrm{grad}}$ terms, whereas the 3D setup captures the dynamics along all three directions leading to a higher domain wall energy. 
This increased energy in 3D is minimized at the cost of higher polarization to bring the FE layer in a steady-state multidomain and results in an overall higher charge response of the MFIM stack (Fig.~\ref{fig:MFIM} (d)).  
Due to similar reasons, the applied voltage required for the multidomain/single-domain transition is higher for the denser domains in 3D.
It is worth noting that the 2D results are in excellent agreement with those reported by Saha and Gupta \cite{saha2020multi}.

\subsection{Domain dynamics and NC effect in 3D MFISM stack}

\begin{figure*}
    \centering
    \includegraphics[width=0.98\linewidth]{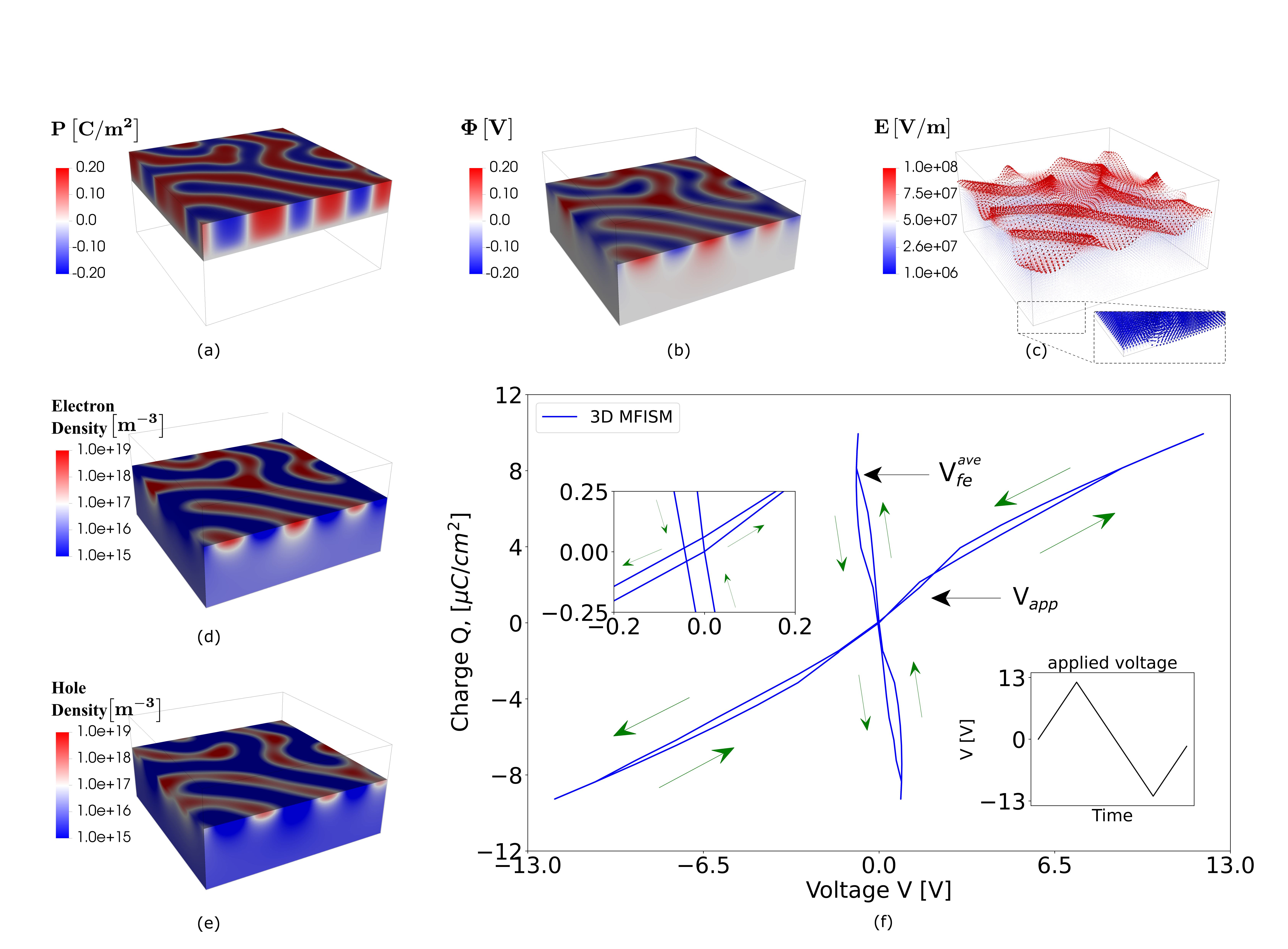}
    \caption{MFISM stack with 5 nm thick HZO on top, followed by 1 nm thick SiO$_2$, and a 10 nm thick Si as the ferroelectric, dielectric, and semiconductor layers, respectively. Vertical direction represents the thickness of the device (z). For an applied voltage, $V_{\rm app} = 0~\rm{V}$ (a) Polarization distribution showing multi-domain formation in FE (b) Potential distribution induced in the semiconductor (c) Electric field vector plot in semiconductor (d,e) electron and hole density respectively in the semiconductor region. (f) $Q-V$ characteristics of the device for varying applied voltage $V_{\rm app}$.}
    \label{fig:MFIS}
\end{figure*}

Now we move our focus to a gate stack of a FeFET which consists of a 5 nm thick HZO, 1 nm thick SiO$_2$, and a 10 nm thick Si as the ferroelectric, dielectric, and semiconductor layers respectively. The purpose of these simulations is to demonstrate the capability of 3D FerroX simulations to study heterostructures governed by the coupling between all three physical processes. The simulation setup is identical to the MFIM case discussed in the previous section with the following two exceptions : (a) the dielectric layer is 1 nm thick SiO$_2(\epsilon_{\rm{SiO}_2} = 3.9)$ instead of 4 nm thick Al$_2$O$_3(\epsilon_{\rm{Al}_2\rm{O}_3} = 10)$ and (b) a 10 nm thick undoped silicon is used as the semiconductor layer. 
In this case, the electron and hole densities are calculated self-consistently with the evolving polarization, following equation (\ref{eq:rho}).
We refer interested readers to ``\nameref{subsec:rho}'' in the Method section for more details. 

The simulation results with $\rm{Al}_2\rm{O}_3$ as DE show reduced electron/hole density, potential, and charge compared to the SiO$_2$ cases and are shown in the supplementary material.
The underlying physics of multidomain formation in FE is the same as in the MFIM and is corroborated by the steady-state multidomain pattern shown in Fig.~\ref{fig:MFIS}(a). 
Similar to the MFIM case described in the previous section, inhomogeneous in-plane potential $\Phi$ and electric field $E$ distributions are formed in the device with the maximum amplitude at the FE-DE interface, shown in Fig.~\ref{fig:MFIS}(b) and (c), respectively. 
This inhomogeneity in the FE-DE interface potential manifests into a spatially varying in-plane potential at the DE-Si interface Fig.~\ref{fig:MFIS}(b). 
This leads to a local accumulation of electrons and holes in an un-doped semiconductor, shown in Fig.~\ref{fig:MFIS}(d) and \ref{fig:MFIS}(e) respectively, exhibiting maxima and minima which correspond to negative and positive $\Phi$ in Fig~\ref{fig:MFIS}(b). For instance, with $V_{\rm app} = 0~\rm{V}$, the minimum DE-Si interface potential is close to $-0.2~\rm{V}$ for MFISM compared to $0~\rm{V}$ for the conventional MIS case (see supplementary materials Fig.~2(b)).
Further exploration is needed to characterize the channel current under the effect of multidomain FE in FeFETs and NCFETs by including electron transport model.



\section*{Discussion}
This work presents a ferroelectric-based transistor simulation framework that is scalable on next-generation GPU/multicore systems.
We present several 3D case studies to demonstrate the accuracy and capability of our methodology.
Specifically, we focused on stacks consisting of ferroelectric materials, dielectric materials, and semiconductors, which is the essential functioning block in FeFETs and NCFETs. 
The investigation of polarization domain dynamics in response to varying applied voltage has revealed a domain-wall induced negative capacitance effect, which is strengthened by denser domain patterns. 
The capability of this solver to scale to at least hundreds of GPUs allows users to comprehensively explore device performance, with both physical and geometrical details addressed, which was difficult to realize prior to this study.
Overall, this work lays the groundwork for future development of exascale-enabled device simulators, as well as adds to the growing body of research on phase-field modeling.
Although this work has successfully demonstrated the core functionality of the simulator, the generalisability of the involved physical mechanisms is subject to certain limitations. 
For instance, only out-of-plane polarization is considered, which leads to underestimated total free energy in the FE layer. Therefore, a natural progression of this work is to include the in-plane polarization components. 
Additionally, strain fields can contribute significantly to the polarization dynamics, which is associated with lattice structure changes \cite{chen2008phase, ashraf2012phase, li2001phase, li2005ferroelectric, li2006temperature, li2008influence}. 
Further code development needs to be conducted to introduce strain into the total free energy. 
Other possible improvements are to include the temperature dependence on the FE domain dynamics \cite{chen2008phase,li2001phase, li2005ferroelectric, li2006temperature, li2008influence, Zubko2016}, and upgrade the electron transport model from the macroscopic Fermi-Dirac distribution function to microscopic transport models describing the evolution of the distribution function, such as the Boltzmann electron transport model.
In the meantime, we are taking active efforts to complete the setup of a full-3D FeFET including the source and drain electrodes, as well as non-planar transistor structures such as finFETs \cite{Krivokapic2017, Seo2018, Choi2001,Yu2002}. 
We will also develop algorithmic improvements, including accelerated schemes for self-consistently solving the Poisson equation, and adaptive mesh refinement methods. 
Altogether, FerroX will provide an increasingly powerful modeling capability for the exploration and design of ferroelectric based transistor devices.

\section*{Methods}
\subsection{Mathematical model}\label{sec:model}
In this section we describe the formulation of the model to simulate devices comprising of ferroelectric, dielectric, and semiconductor thin films stacked between two metal plates as shown in Fig.~\ref{fig:overview}. 
In order to model the interaction between these layers, we self-consistently solve the time-dependent Ginzburg Landau (TDGL) equation for ferroelectric polarization, Poisson's equation for electric potential, and semiconductor charge equation for carrier densities in semiconductor regions.

\subsection{Ginzburg-Landau model for ferroelectrics}\label{subsec:TDGL}
The total free energy $F$ of the system in the TDGL equation is described as  
\begin{equation}
    F = \int_V f(\mathbf{r})dV
\label{eq:F}
\end{equation}
where the total energy density $f(\mathbf{r})$ of the system includes the bulk Landau free energy $(f_{\rm Land})$, the domain wall energy or gradient energy $(f_{\rm grad})$, and the electric energy of the applied electric field ($f_{\rm elec}$)~\cite{chen2008phase,chen1998applications,saha2019phase,saha2020multi,ashraf2012phase}
\begin{equation}
    f(\mathbf{r}) = f_{\rm Land}(\mathbf{r}) + f_{\rm grad}(\mathbf{r}) + f_{\rm elec}(\mathbf{r}).     
\label{eq:f}
\end{equation}
The phenomenological formalism~\cite{devonshire1949xcvi} for the internal free energy $f_{\rm Land}$ is  given by the well-known  Landau–Ginzburg–Devonshire (LGD) polynomial form as a function of the spontaneous polarization $P$ as 
\begin{equation}
    f_{\rm Land} = \frac{1}{2}\alpha P^2 + \frac{1}{4}\beta P^4 + \frac{1}{6}\gamma P^6
\label{eq:free}
\end{equation}
where $\alpha$, $\beta$, and $\gamma$ are Landau free energy coefficients. For ferroelectrics, $\alpha$ must be negative and $\gamma$ is positive, which gives rise to double-well energy landscape leading to energy minima at $\pm P_\mathrm{remnant}$ . This implies that the  material is unstable at the depoled state $(P = 0)$, so its response to the external voltage is opposite to that of normal dielectrics, which leads to the origin of the NC effect in ferroelectric thin films~\cite{salahuddin2008use}. 

In addition to the free energy component, spatial variations in polarization contribute to gradient energy which characterizes the energy from dipole–dipole interactions resulting from
spatially inhomogeneous polarization~\cite{chen2008phase,chen1998applications,saha2019phase,saha2020multi,ashraf2012phase}. Gradient energy density, $f_{\rm grad}$ can be expressed as
\begin{equation}
    f_{\rm grad} = \frac{1}{2}\left[g_{44}\left(\frac{\partial P(\mathbf{r})}{\partial x}\right)^2 + g_{44}\left(\frac{\partial P(\mathbf{r})}{\partial y}\right)^2 + g_{11}\left(\frac{\partial P(\mathbf{r})}{\partial z}\right)^2\right]
\label{eq:grad}
\end{equation}
where $g_{11}$ and $g_{44}$ are gradient energy coefficients. 

The electrostatic energy density $(f_{\rm elec})$ is given by~\cite{chen2008phase,chen1998applications,saha2019phase,saha2020multi,ashraf2012phase}
\begin{equation}
    f_{\rm elec} = \mathbf{E}\cdot\mathbf{P} = E_z\cdot P
\label{eq:elec}
\end{equation}
where $E_z = -\frac{d\Phi}{dz}$, with $\Phi$ as the electric potential distribution in the device and is obtained by solving Poisson's equation as discussed next.

Using equations (\ref{eq:F})-(\ref{eq:elec}) in equation (\ref{eq:TDGL1}), we obtain the TDGL equation as shown below:
\begin{equation}
    \frac{-1}{\Gamma}\frac{\partial P(\mathbf{r},t)}{\partial{t}} = \alpha P + \beta P^3 + \gamma P^5 - g_{44}\frac{\partial^2P}{\partial x^2} - g_{44}\frac{\partial^2P}{\partial y^2} - g_{11}\frac{\partial^2P}{\partial z^2} + \frac{d\Phi}{dz}
\label{eq:TDGL2}
\end{equation}
\subsection{Poisson's equation for electric potential}
Distribution of electric potential in these systems is obtained by solving Poisson's equation in the following form:

\begin{equation}
    \nabla\cdot\epsilon\nabla\Phi = \frac{\partial P}{\partial z} - \rho(\Phi)\label{eq:Poisson}
\end{equation}
where $\epsilon$ is a spatially-varying permittivity and $\rho$ is the total free charge density in the semiconductor region and can be set to zero for MFIM devices because semiconductor is not present.

\subsection{Charge density in semiconductor region}\label{subsec:rho}
The total charge density in the semiconductor region can be calculated from carrier (electron and hole) densities using equation (\ref{eq:rho}) which are estimated using Fermi-Dirac distribution function. For example, the equilibrium concentration of electrons per unit volume in a three-dimensional semiconductor can be estimated as~\cite{neamen2003semiconductor}
\begin{equation}
    n_e(\mathbf{r}) = \int_{E_c}^\infty g(E)f_{FD}(E)dE = \int_{E_c}^\infty \frac{g(E)}{1 + e^{(E - e\Phi)/{k_BT}}}dE    
\label{FD}
\end{equation}
where $g(E)$ is the density of states, $f_{FD}(E)$ is the Fermi-Dirac distribution function, $k_B$ is the Boltzmann constant, $T$ is the temperature, and $E_c$ is energy at the edge of the conduction band. The density of states for electrons is given by
\begin{equation}
    g(E) = \frac{(2m^*_e)^{3/2}}{2\pi^2\hbar^3}\sqrt{E - E_c}
\label{eq:density_state}
\end{equation}
where $m_e^*$ is the effective mass of electrons and $\hbar$ is the reduced Planck's constant. Substituting equation (\ref{eq:density_state}) in equation (\ref{FD}), we obtain the electron density as 
\begin{equation}
    n_e(\mathbf{r}) = \frac{(2m^*_e)^{3/2}}{2\pi^2\hbar^3}\int_{E_c}^\infty \frac{\sqrt{E - E_c}}{1 + e^{(E - e\Phi)/{k_BT}}}dE.
\label{eq:ne}
\end{equation}
Density of holes can be estimated in a similar fashion and can be expressed as
\begin{equation}
    n_p(\mathbf{r}) = \frac{(2m^*_p)^{3/2}}{2\pi^2\hbar^3}\int_{-\infty}^{E_v} \sqrt{E_v - E}\left[ 1 - \frac{1}{1 + e^{(E - e\Phi)/{k_BT}}}\right]dE
\label{eq:np}
\end{equation}
where $m_p^*$ is the effective mass of holes and $E_v$ is the edge of the valence band. The Fermi-Dirac distribution approaches the Maxwell-Boltzmann distribution in the limit of high temperature and low particle density. Therefore, if we assume $E - e\Phi >> k_BT$ and $E - e\Phi << -k_BT$ in the expressions for $n_e$ and $n_p$ we obtain the simple Maxwellian distribution for the density of electrons and holes in the following form:
\begin{eqnarray}
    n_e(\mathbf{r}) &= N_ce^{-(E_c - e\Phi)/{k_BT}} \nonumber\\
     n_p(\mathbf{r}) &= N_ve^{-(e\Phi - E_v)/{k_BT}}
     \label{MaxBolt_ne_np}
\end{eqnarray}
where $N_c$ and $N_v$ are the effective densities of state of conduction and valance band respectively.

\subsection{Polarization boundary conditions}
Due to the symmetry breaking on the surface of ferroelectrics, polarization on surfaces is different from that inside the crystal. In the study of ferroelectric thin films, the surface effect can be included in continuum theories by setting appropriate boundary conditions of polarization. Such a \textbf{boundary condition with surface effect} for polarization, $P$, along the thickness ($z-$ direction) of the film is usually given by~\cite{saha2020multi,saha2019phase}
\begin{equation}
    \lambda \frac{dP}{dz} - P = 0
\label{Pbc}
\end{equation}
where $\lambda$ is the so-called extrapolation length, which is an artificial parameter introduced to describe the difference of polarizations between the surface and the interior of the material. The polarization is reduced at the surface when $\lambda$ is positive or zero, while it is enhanced at the surface when $\lambda$ is negative. When $\lambda$ approaches infinity, the boundary condition
becomes 
\begin{equation}
    \frac{dP}{dz} = 0,    
\label{PbcFree}
\end{equation}
the so-called \textbf{free polarization boundary condition}~\cite{wang2010effect}, which means that there is no difference in polarizations in the media between the surface and the interior. When $\lambda$ equals zero, polarization is completely suppressed at the surface, i.e.,
\begin{equation}
    P = 0,    
\label{PbcDir}
\end{equation}
which is the \textbf{zero polarization boundary condition}~\cite{wang2010effect}. 
We have implemented all three boundary conditions for the user to specify the appropriate polarization boundary condition and specify a value for $\lambda$ at run time. Along the thickness of the film, boundary conditions are set at the metal-ferroelectric and ferroelectric-dielectric interfaces. We calculate $\frac{dP}{dz}$ and $\frac{d^2P}{dz^2}$ using either the values of $P$ or $\frac{dP}{dz}$ at these interfaces using a second order accurate, one-sided, three-point stencil. 
In this work we enforce periodic boundary conditions in the planar directions.
Further implementation of boundary conditions with surface effects are our ongoing efforts. 

\subsection{Electrical boundary conditions}
Electrical boundary conditions are needed to enforce the continuity of the normal component of the displacement vector throughout the computational domain and are controlled by specifying boundary conditions for the Poisson solver. We solve Poisson's equation for single state variable, i.e. the potential distribution $(\Phi)$ in the entire computational domain. Therefore, boundary conditions are needed only on the surfaces of the device. We assume \textbf{periodic} boundary conditions along the transverse directions $x$ and $y$, and use a \textbf{Dirichlet} boundary condition to mimic the applied voltage $V_{\rm app}$ along the thickness of the device through the metal contacts as $\Phi(x,y,z_{min}) = V_0$ and $\Phi(x,y,z_{max}) = V_1$, where $V_{\rm app} = V_1 - V_0$.
\subsection{Numerical Approach and Implementation}\label{sec:numerical}
In this section we describe the numerical algorithm to solve the system of equations governing the dynamics of the system. The two fundamental kernels consist of an integration of $P$, and the solution of a Poisson solve to compute $\Phi$.
The integration of $P$ is performed using equation (\ref{eq:TDGL2}), which we write compactly as:
\begin{equation}
    \frac{\partial P}{\partial t} = f(P,\Phi)
    \label{TDGL_simp}
\end{equation}
We have implemented a first and second-order temporal integrator.
We utilize the Multi-Level-Multi-Grid (MLMG) geometric multigrid linear solver built in the AMReX library~\cite{zhang2021amrex} to solve Poisson's equation. It is a second order accurate iterative solver which achieves convergence by progressively minimizing the residual below an user-defined threshold. Periodic boundary conditions are set in the two in-plane directions, $x$ and $y$, while a Dirichlet boundary condition is used along the thickness direction $(z)$ of the device to control the applied voltages. The right-hand-side of the Poisson's equation also depends on the charge density $(\rho)$, which is non-zero only in the semiconductor region. The integrals in equation (\ref{eq:ne}) and (\ref{eq:np}) are computed using an  approximation to the Fermi-Dirac integral of order 1/2 \cite{bednarczyk1978approximation} and then equation (\ref{eq:rho}) is used to estimate the total charge density.

The solution of the Poisson problem is not as straightforward for problems with a semiconductor layer, since right-hand-side of the Poisson solve is also a function of $\Phi$.  In this case, we use a fixed-point strategy where we iteratively lag the effects of $\Phi$ when computing the right-hand-side, and iterate to convergence.  We stop the iterations when the average magnitude of change over all cells is less than a user-defined tolerance.  In our example, we use a tolerance of $10^{-5}$ and observe that the solution converges in typically 2 iterations.

In our algorithm, the superscript denotes the time level, i.e., $(P,\Phi)^n$ denotes the solution at $t^n \equiv n\Delta t$.
The overall numerical scheme proceeds as follows:\\ \\
{\bf INITIALIZATION:} Given $P^0$, compute $\Phi^0$ using the following:
\begin{itemize}
    \item {\bf Step 0:} Define $\Phi^{0,(0)} = 0$ and then iterate equation (\ref{eq:iterate0}) over $k=0,\cdots$ until the desired tolerance is achieved.  Again, note that iterations are only required if there is a semiconductor layer where $\rho$ depends on $\Phi$.
    \begin{equation}
        \nabla\cdot\epsilon\nabla\Phi^{0,(k+1)} = \frac{\partial P^0}{\partial z} - \rho(\Phi^{0,(k)})\label{eq:iterate0}
    \end{equation}
    Then, set $\Phi^0 = \Phi^{0,(k+1)}$.
\end{itemize}

{\bf TIME-ADVANCEMENT:} Given $(P,\Phi)^n$, compute $(P,\Phi)^{n+1}$ using the following steps:\\ \\
{\bf Predictor}
\begin{itemize}
    \item {\bf Step 1a:} Compute $P^{n+1,*} = P^n + \Delta t f(P^n,\Phi^n)$.
    \item {\bf Step 1b:} Define $\Phi^{n+1,*,(0)} = \Phi^n$ and then iterate equation (\ref{eq:iterate1}) over $k=0,\cdots$ until the desired tolerance is achieved:
    \begin{equation}
        \nabla\cdot\epsilon\nabla\Phi^{n+1,*,(k+1)} = \frac{\partial P^{n+1,*}}{\partial z} - \rho(\Phi^{n+1,*,(k)})\label{eq:iterate1}
    \end{equation}
    Then, set $\Phi^{n+1,*} = \Phi^{n+1,*,(k+1)}$.
\end{itemize}
If only first-order accuracy in time is desired, the corrector step can be skipped.\\ \\
{\bf Corrector}
\begin{itemize}
    \item {\bf Step 2a:} Compute $P^{n+1} = P^n + \frac{\Delta t}{2} f(P^n,\Phi^n) + \frac{\Delta t}{2} f(P^{n+1,*},\Phi^{n+1,*})$.
    \item {\bf Step 2b:} Define $\Phi^{n+1,(0)} = \Phi^{n+1,*}$ and then iterate equation (\ref{eq:iterate2}) over $k=0,\cdots$ until the desired tolerance is achieved:
    \begin{equation}
        \nabla\cdot\epsilon\nabla\Phi^{n+1,(k+1)} = \frac{\partial P^{n+1}}{\partial z} - \rho(\Phi^{n+1,(k)})\label{eq:iterate2}
    \end{equation}
    Then, set $\Phi^{n+1} = \Phi^{n+1,(k+1)}$.
\end{itemize}

\subsection{Implementation}
We implement our code using the AMReX software library \cite{zhang2021amrex}, which is developed and supported by the DOE Exascale Computing Project Block-Structured Adaptive Mesh Refinement Co-Design Center. AMReX contains many features for solving partial differential equations on structured grids; here we discuss relevant features for our present implementation, as well as future plans that will incorporate additional features. AMReX manages data and operations on structured grids in a manner that can efficiently use the full range of computer architectures from laptops to manycore/GPU supercomputing architectures. We divide the computational domain into non-overlapping grids, and each grid is assigned to an MPI rank. AMReX uses an MPI+X finegrained parallelization strategy, where X can be OpenMP (for multicore architectures), or CUDA (for GPU-based architectures). Each of these strategies are implemented with the same front-end code using specialized looping structures within AMReX and the portability across various platforms is ensured by AMReX during compile-time. Each MPI process applies computational kernels only to the data in grids that they own in the form of triply-nested loops (over each spatial dimension). For pure MPI calculations, the loop is interpreted as a standard “i/j/k” loop. For MPI+OpenMP calculations, the bounds of the loop is further subdivided over logical tiles, and each OpenMP thread loops over a particular tile. For MPI+CUDA calculations, AMReX performs a kernel launch and offloads each data point to CUDA threads that in turn perform computations. AMReX manages data movement by keeping data on the GPU devices as much as possible, avoiding costly communication between the host and device. Thus, whenever possible, data movement to/from the host/GPU and also between MPI ranks is limited to ghost cell data exchanges, which occur a small number of times per time step (excluding the linear solvers). In Section \ref{sec:scaling} we demonstrate the efficiency and scalability of our code using pure MPI and MPI+CUDA simulations on NERSC systems. This analysis is realized by built-in AMReX profiling tools; however more in-depth analysis is possible with an extensive variety of compatible profilers such as CrayPat, IPM, and Nsight. Data from the simulation can be efficiently written using a user-defined number of MPI ranks, to prevent overwhelming the system with simultaneous writes. Visualization can be performed with a number of publicly-available software packages, including Amrvis, VisIt, Paraview, and yt.

\subsection{Numerical Validation}\label{sec:numericalvalidation}
In this Section we present the numerical convergence of the algorithm in time and space by computing the reduction in error as we increase the temporal and/or spatial resolution.

Our temporal integrators are designed to produce first or second-order convergence in time.
Our spatial discretization in second-order within each material, however discontinuities in our model near interface boundaries will result in a reduction to first-order in space.
For example, the abrupt change in the dielectric coefficients at interfaces results in order reduction of the electric potential from the Poisson solver.
Also, the abrupt change in the right-hand-side of the Poisson equation from $\nabla\cdot{\bf P}$ (in the ferroelectric region), to zero (in the dielectric region), to the charge density (in the semi-conductor region) will also result in spatial order reduction.

The physical problem setup is as follows:
the three-dimensional domain is 32~nm on each side.
The lowest 25\% of the domain is semiconductor material, the next 25\% is dielectric, and the upper 50\% of the domain is ferroelectric. Material parameters listed in Table \ref{tab:sim_param} are utilized in all these tests. 
We set $\Phi=0$ on the upper and lower domain boundaries in $z$ and periodic boundary conditions in $x$ and $y$.
We use the boundary condition with surface effects for $P$ given by (\ref{Pbc}) with $\lambda = 3$~nm.
The initial condition for polarization is a smooth Gaussian bump isolated to the ferroelectric region, given by
\begin{equation}
    P = 0.002 e^{-\left(\frac{x^2}{2\sigma_1^2} + \frac{y^2}{2\sigma_1^2} + \frac{(z-z_0)^2}{2\sigma_2^2}\right)},
\end{equation}
with $\sigma_1=5$~nm, $\sigma_2=2$~nm, and $z_0=24$~nm.

In the first test, we measure temporal order of accuracy by performing a series of simulations on a fixed grid structure, but with three different time steps using the first-order temporal scheme.
Thus, the notion of ``coarse'', ``medium'', and ``fine'' simulations are referring to temporal resolution.
In the second test, we repeat this procedure but use the second-order temporal scheme.
For the first two tests, we use $128^3$ grid cells (0.25~nm resolution) with $\Delta t=50, 25$, and $12.5$~fs.
In the third test, we measure spatial order of accuracy by performing a series of simulations using the same time step, but with three different spatial resolutions.
For this third test, the notion of ``coarse'', ``medium'', and ``fine'' simulations are referring to spatial resolution
and we use $\Delta t=25$~fs and $64^3, 128^3$, and $256^3$ grid cells (0.5, 0.25, and 0.125~nm resolution).
We run all simulations to a total time of $t=400$~fs.

The convergence rate is defined as the base-2 log of the ratio of errors between the coarse-medium, $E_c^m$, and medium-fine solutions, $E_m^f$.
\begin{equation}
{\rm Rate} = \log_2\left(\frac{E_c^m }{E_m^f}\right)
\end{equation}
 $E_c^m$ is computed as the $L^2$ norm of the difference of the coarse and medium solution.
 For the spatial convergence test, these two solutions are at different resolutions; thus we average the medium solution down to coarse resolution by averaging the 8 medium cells overlying a coarse cell.
 $E_m^f$ is computed in the same way, but with the medium and fine solutions.
Formally, for the scalar potential field $\Phi$ (and same procedure for $P$) this is written as:
\begin{eqnarray}
E_c^m &= \sqrt{\frac{1}{N_{\rm c}} \sum_{i,j,k} \lvert\Phi_{m} - \Phi_{c}\rvert^2} ; \nonumber \\
E_m^f &= \sqrt{\frac{1}{N_{\rm m}} \sum_{i,j,k} \lvert\Phi_{f} - \Phi_{m}\rvert^2}
\end{eqnarray}
where, $\Phi_c$, $\Phi_m$, and $\Phi_f$ are the fields obtained from the coarse, medium, and fine solutions, and $N_{\rm c}$ and  $N_{\rm f}$ are total number of grid points used for the coarse and medium simulations, respectively. 

In Table \ref{tab:convergence} we show convergence results for $P$ and $\Phi$.
For Test 1 we see the expected first-order convergence in time.
For Test 2 we see the expected second-order convergence in time.
For Test 3 we see an expected reduction to first order in the electric potential, and a slight reduction in order for the polarization.
This is expected since there are a number of spatial discontinuities present in the code that should not exhibit second-order convergence, as described above.

\newcommand{\ra}[1]{\renewcommand{\arraystretch}{1.2}}
\begin{table*}\centering
\resizebox{\columnwidth}{!}{
\small
\ra{1.0}
\begin{tabular}{@{}rccccccc@{}}\toprule
& \multicolumn{3}{c}{$P$} & \phantom{abc}& \multicolumn{3}{c}{$\Phi$} \\
\cmidrule{2-4} \cmidrule{6-8} 
& $E_c^m$ & Rate & $E_m^f$ && $E_c^m$ & Rate & $E_m^f$ \\ \midrule
Test 1 & $3.53\times 10^{-8}$ & 1.00 & $1.77\times 10^{-8}$ && $1.12\times 10^{-6}$ & 1.01 & $5.53\times 10^{-7}$ \\
Test 2 & $2.31\times 10^{-10}$ & 2.02 & $5.70\times 10^{-11}$ && $1.79\times 10^{-8}$ & 2.07 & $4.25\times 10^{-9}$ \\
Test 3 & $6.28\times 10^{-7}$ & 1.96 & $1.61\times 10^{-7}$ && $2.32\times 10^{-5}$ & 1.00 & $1.16\times 10^{-5}$ \\
\bottomrule
\end{tabular}
}
\\
\caption{Convergence study for an MFISM test problem using (Test 1) temporal refinement only, first-order temporal scheme, (Test 2) temporal refinement only, second-order temporal scheme, and (Test 3) spatial refinement only, second-order temporal scheme}
\label{tab:convergence}
\end{table*}

\subsection{Code performance}{\label{sec:scaling}}
We asses the performance and scalability of our code on HPC systems with tests on the NERSC supercomputers.
We consider the Haswell partition on the Cori system and NERSC's newest flagship supercomputer, Perlmutter. The Haswell partition consists of 2,388 nodes; each node contains an Intel Xeon “Haswell” Processor with 32 physical cores and 128GB memory. Perlmutter system has 1536 GPU accelerated compute nodes each consisting of 4 NVIDIA A100 GPUs with 160 GB memory per node. We perform weak scaling tests and then compare times across different architectures using pure MPI runs on haswell CPUs and MPI+CUDA runs on Perlmutter GPUs.  To summarize, we achieve near perfect scaling to several hundred GPUs, with a 15x speedup over CPU runs on a node-by-node basis.

We perform the weak scaling tests with the set-up of a MFIM device. We consider the interface between ferroelectric and dielectric in the middle of the computational domain along $z$ (thickness) direction. Transverse directions of the film align with the $x$ and $y$ directions respectively. We perform two sets of simulations on NERSC systems described above. The first set of simulations employ the pure MPI paradigm on the Haswell partition with a maximum of 32 MPI processes per node. The second set of simulations employ the MPI+CUDA parallelization strategy on the GPU accelerated compute nodes of perlmutter, which has a maximum of 4 NVIDIA A100 GPUs per node, and we use one MPI processor per GPU. The baseline case for all weak scaling tests performed with different parallelization strategies has a computational domain size of $32.0 \rm{nm} \times 32.0 \rm{nm} \times 32.0 \rm{nm}$ uniformly discretized using $256\times256\times256$ cells (16,777,216 total). For this base case we use 1/4th of a node in each simulation set, i.e. 8 MPI ranks on haswell partition of cori and 1 MPI rank + 1 GPU on Perlmutter compute node. We should note that this simulation utilizes $\sim 5.5$ GB of memory which is $\sim 14\%$ of total available memory per per Perlmutter GPU. For the weak-scaling study, we increase the domain size and number of cells in each direction consistent with the increase in the core count to maintain a constant amount of computational work per core. Weak scaling results for the two sets of simulations, indicating the average  simulation time per time step as a function of the number of nodes is shown in Fig.~\ref{fig:scaling}. We omit the timing associated with initialization as this occurs only once at the beginning of the simulation. We performed tests on each system up to 128 nodes, which corresponds to 4096 CPUs on haswell and 512 GPUs on Perlmutter. For the 1/4th node baseline runs, the GPU simulations performed 11x faster than the pure MPI simulations on the Haswell partition. We also observe that for the 2-node through 128-node runs, the GPU simulations performed approximately 15x faster than the pure MPI simulations on the Haswell partition. Both the Haswell partition and Perlmutter GPU simulations achieve nearly perfect scaling up to 128 nodes beyond the 1-node threshold. This is attributed to the fact that as we increase the number of processors within the node, the MPI communication time keeps increasing until the node is saturated and then the computation-to-communication ratio remains very similar across all nodes for the problem sizes we consider.


\begin{centering}
\section*{Supplementary Material}
\end{centering}

\section*{P-E characteristics of 10 nm HZO thin-film}
We studied the P-E characteristics of a 10 nm HZO thin-film using 3D FerroX simulation and compared the results with experimental measurements~\upcite{si2019ultrafast}. For this, we simulate an MFM device consisting of a 10 nm thick HZO film between two metal plates with an in-plane dimension of 16 nm $\times$ 16 nm. Polarization is initialized as uniformly distributed random number in $[-0.002,0.002]$ in the FE and the applied voltage across the device is varied by specifying the Dirichlet boundary condition to the Poisson solver at the metal contacts. Physical and numerical parameters same as those shown in Table 1 of the main manuscript are used. 
\begin{figure}[h]
    \centering
    \includegraphics[width=0.6\linewidth]{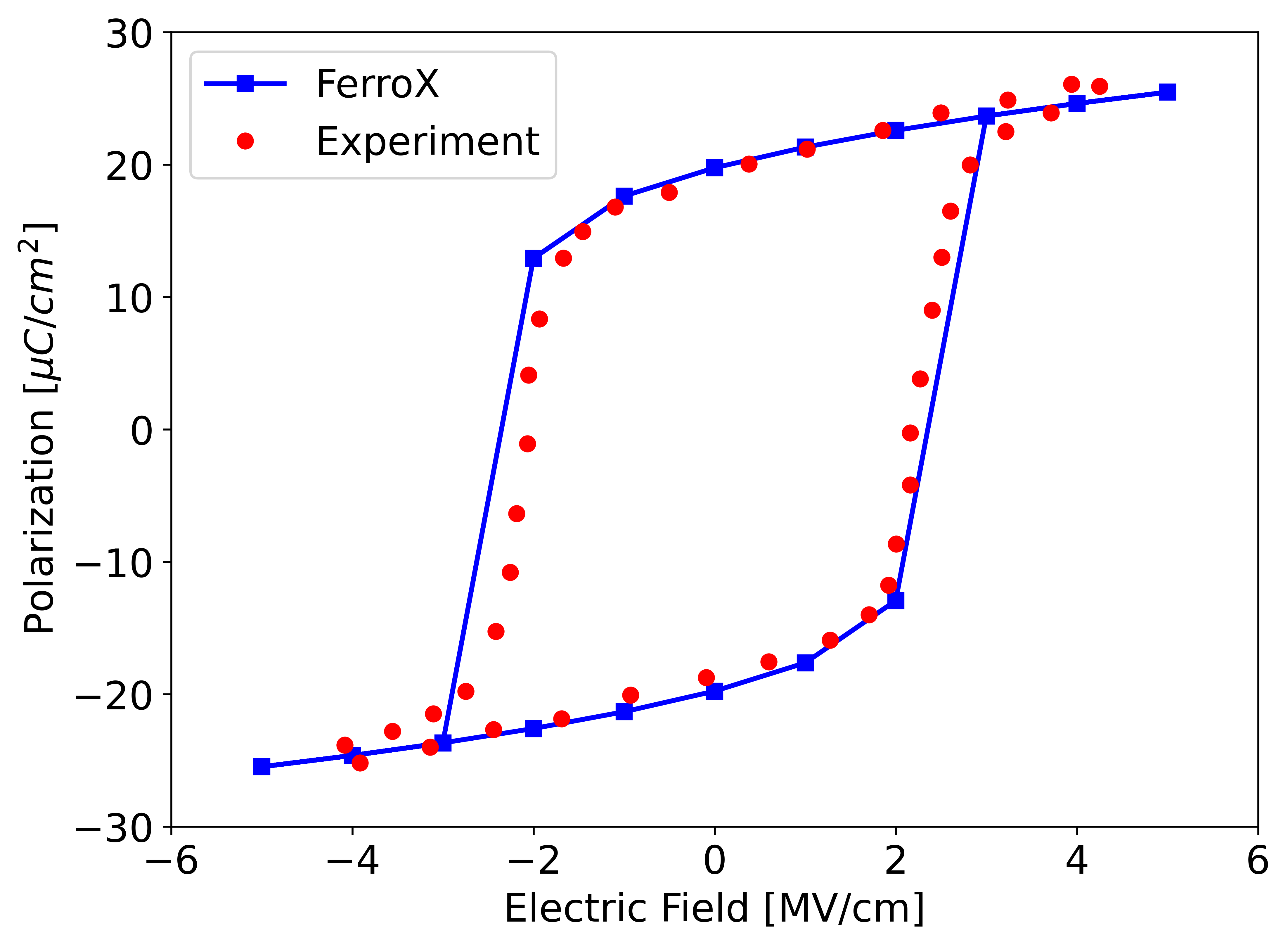}
    \caption{P-E characteristics obtained using 3D FerroX simulation of an MFM device with 10 nm HZO film as the FE. Spatially averaged polarization is plotted against the applied electric field.}
    \label{fig:MFM_PE}
\end{figure}
Tha applied voltage is varied from -5.0 to 5.0 V and for each applied voltage, the FE polarization is allowed to evolve until a steady state is reached. We plot the spatially averaged polarization in the FE against the electric field to obtain the P-E characteristics shown in Fig.~\ref{fig:MFM_PE}. Simulation results are in excellent agreement with the experimental measurements~\cite{si2019ultrafast}, thus validating the implementation of TDGL and Poisson's equation in FerroX.

\section*{2D Simulation of MFISM device}
In this section, we study multi-domain formation in a 2D MFISM device using FerroX and compare our results with previous 2D work~\upcite{saha2020multi}. This test case serves as a verification of the algorithms implemented in FerroX. All the physical and numerical parameters are identical to the 3D case presented in the main manuscript except that the $y-$dimension is not modeled, ignoring the dynamics along that direction. An applied voltage of $V_{\rm app} = 0~\rm{V}$ is used. 
\begin{figure}[h]
    \centering
    \includegraphics[width=\linewidth]{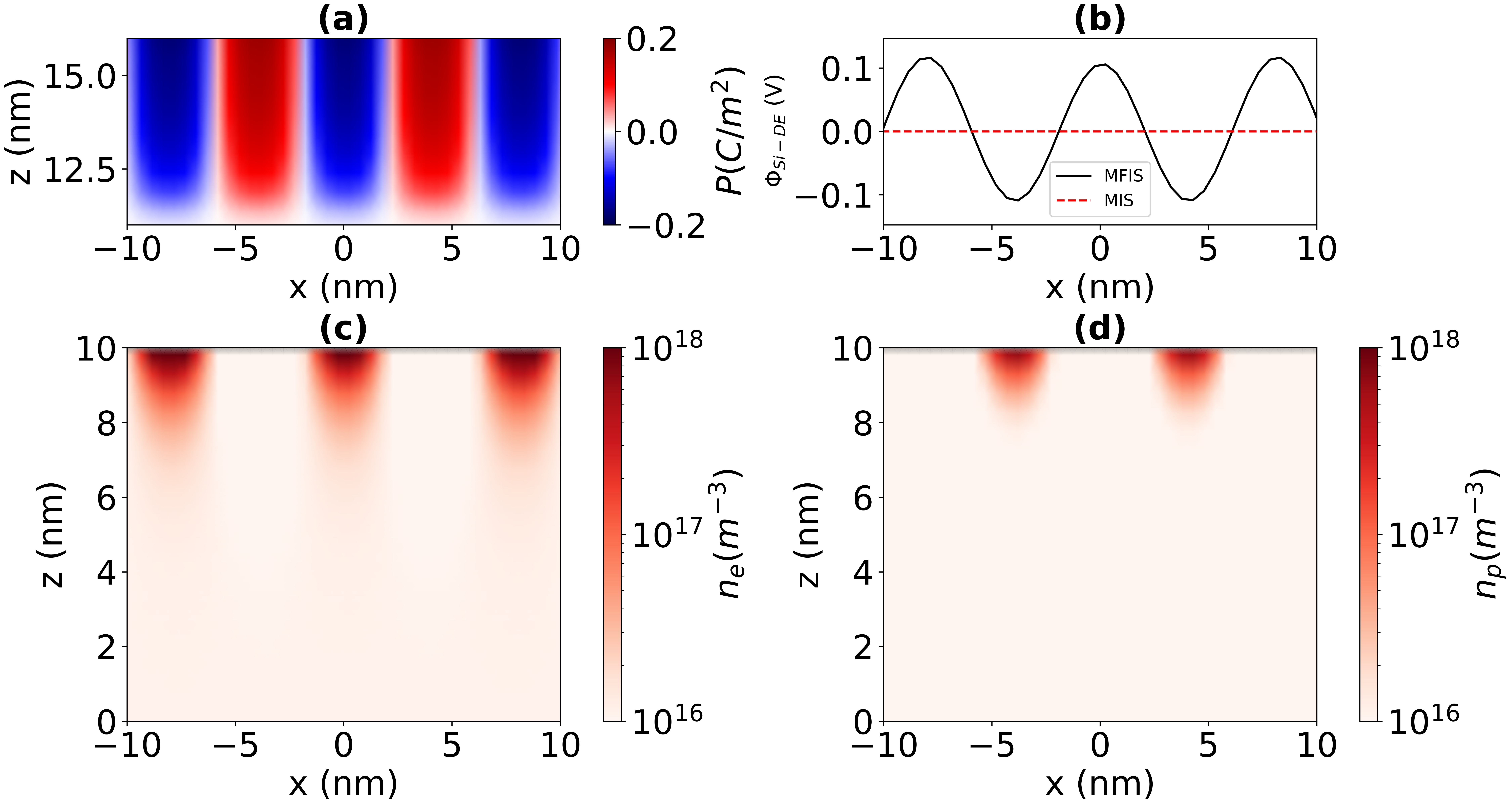}
    \caption{2D FerroX simulation of an MFISM stack. 5 nm thick HZO, 1 nm thick SiO2, and a 10 nm thick Si as the ferroelectric, dielectric, and semiconductor layers respectively using an applied voltage, Vapp = 0 V. (a) Polarization distribution showing multi-domain formation in FE (b) surface potential at FE-Si interface with and without FE (c) and (d) electron and hole density respectively in the semiconductor region.}
    \label{Supp:MFIS}
\end{figure}
Simulation results are shown in Figure \ref{Supp:MFIS}. Similar to the 3D case, the depolarization field, generated at the HZO-SiO$_2$ interface, which acts opposite to the polarization and results in an increase in the depolarization energy is compensated by the formation of $180^\circ$ domains of alternate positive and negative $P$ values of approximately equal size as shown in Fig. \ref{Supp:MFIS}(a). The non-homogeneity in the FE-DE interface potential manifests into a spatially varying potential profile at the DE-Si interface which shows maxima and minima corresponding to negative and positive P values as shown in Fig \ref{Supp:MFIS}(b). Electron and hole density distribution are shown in  \ref{Supp:MFIS}(c) and (d). These results are in good agreement with Figure 6(a),(b), and (c) of Saha et.al.~\cite{saha2020multi}.  

\section*{MFISM simulation with Al$_2$O$_3$ as the dielectric}
Here we study the effect of replacing 1 nm SiO$_2$ with a 4 nm Al$_2$O$_3$ as the dielectric layer in the MFISM stack. 3D simulation with all parameters identical to the 3D MFISM result described in the main manuscript were performed. 
\begin{figure}[h]
    \centering
    \includegraphics[width=\linewidth]{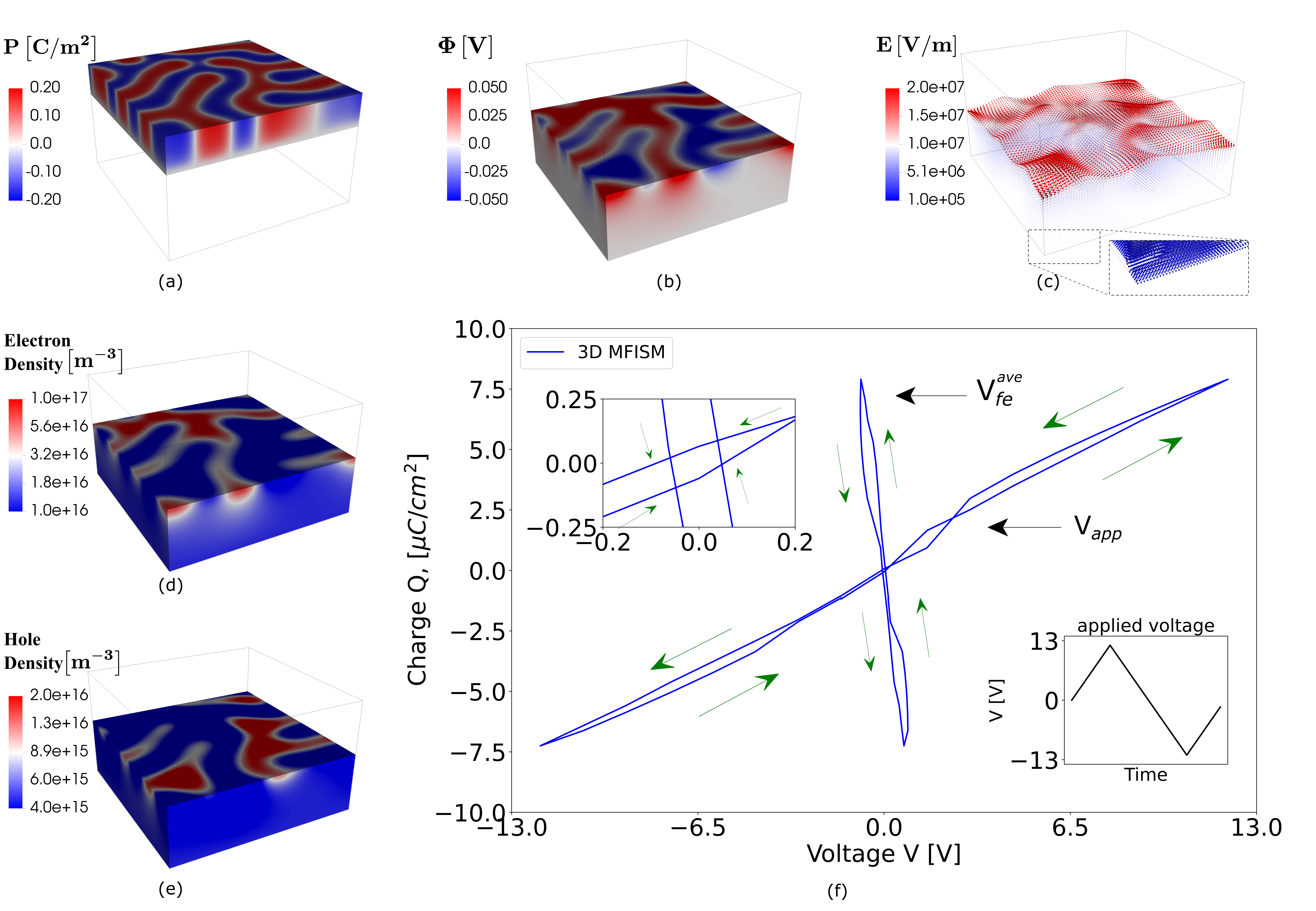}
    \caption{MFISM stack with 5 nm thick HZO on top, followed by 4 nm thick Al$_2$O$_3$, and a 10 nm thick Si as the ferroelectric, dielectric, and semiconductor layers respectively. Vertical direction represents the thickness of the device (z). For an applied voltage, $V_{\rm app} = 0~\rm{V}$ (a) Polarization distribution showing multi-domain formation in FE (b) Potential distribution induced in the semiconductor (c) Electric field vector plot in semiconductor (d,e) electron and hole density respectively in the semiconductor region. (f) Q-V characteristics of the device.}
    \label{Supp:MFISM_Al}
\end{figure}
Results are shown in Fig. \ref{Supp:MFISM_Al}. The differences between these results and the ones obtained using SiO$_2$ arise from the thickness of the dielectric (4 nm in this case vs 1 nm in the case of SiO$_2$) and the relative permittivity (10 vs 3.9 in the case of SiO$_2$) in the dielectric region. Since the potential is maximum at the FE-DE interface, and it decreases as z decreases (away from the interface and towards the semiconductor), the value of potential in the semiconductor region (shown in Fig.  \ref{Supp:MFISM_Al}(b)) is smaller in magnitude by roughly a factor of 4. Corresponding electric field (Fig. \ref{Supp:MFISM_Al}(c)), electron density (Fig. \ref{Supp:MFISM_Al}(d)), and hole density (Fig. \ref{Supp:MFISM_Al}(e)) profiles also show a reduced magnitude. $Q-V$ characteristics shown in Fig. \ref{Supp:MFISM_Al}(f) indicates that the overall charge across the device and average charge in the FE is smaller in this case.  

\section*{Effect of changing dielectric on $Q-V_{fe}$ characteristics}
\begin{figure}[h]
    \centering
    \includegraphics[width=0.9\linewidth]{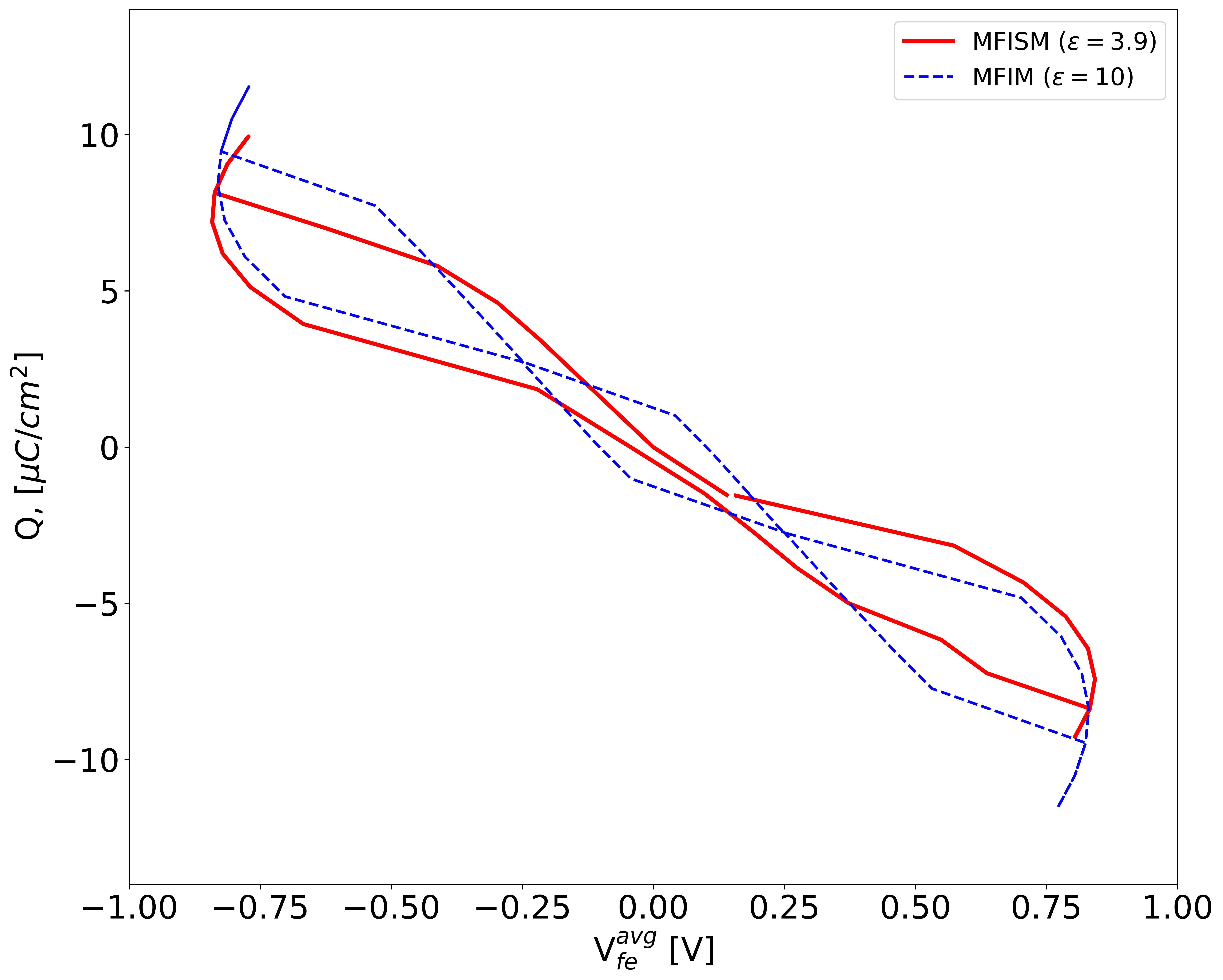}
    \caption{Effect of changing dielectric on $Q-V_{fe}$ characteristics: MFIM results are obtained using Al$_2$O$_3 (\epsilon_{x(y)} = 10)$ as the dielectric whereas MFISM results are obtained using SiO$_2 (\epsilon_{x(y)} = 3.9)$. The MFISM curve has a smaller slope indicating larger negative capacitance effect when dielectric permittivity is smaller.}
    \label{Supp:epsilon_change}
\end{figure}
The amplitude of polarization and the width of the positive and negative domains are similar to the 3D MFIM case. 
However, the $Q-V_{fe}^{avg}$ characteristics (Fig. \ref{Supp:epsilon_change}(f)) shows a higher negative capacitance compared to the MFIM case.
This can be explained by the decrease in the value of the permittivity of the dielectric layer \upcite{saha2020multi}. 
Fig. \ref{Supp:epsilon_change}(c) shows the electric field vector plot which indicates the presence of large in-plane $(x$ and $y$ components$)$ of the field. 
Compared to the MFIM case, as $\epsilon_{x(y),DE}$ decreases, $E_{x(y),DE}$ increases. 
Since the in-plane electric field is continuous across the FE-DE interface, $E_{x(y),FE}$ also increases and results in a higher electrostatic energy density in the ferroelectric. 
Following the analysis in the MFIM case, we conclude that this will lead to an increase in the negative capacitance evident in the higher slope of the $Q-V_{\rm fe}^{\rm avg}$ curve in Fig. \ref{Supp:epsilon_change}(f). 

In the analysis presented in section entitled ``Multi-domain dynamics in MFISM stack", we have studied the effect of changing the dielectric on $Q-V_{fe}$ characteristics. MFISM set up uses SiO$_2 (\epsilon_{x(y)} = 3.9)$ instead of Al$_2$O$_3 (\epsilon_{x(y)} = 10)$, as the dielectric used in MFIM. Figure~\ref{Supp:epsilon_change} shows the comparison of $Q-V_{fe}$ characteristics for these two cases. As discussed in the main manuscript, negative capacitance increases with a decrease in the  permittivity of the dielectric layer.
\section*{Data availability}
The datasets generated during and/or analysed during the current study are available in the zenodo repository at \url{https://doi.org/10.5281/zenodo.7221895}~\cite{kumar_prabhat_2022_7221895}.

\section*{Code availability}
All codes are publicly available on the open-source github platform. The AMReX library can be obtained at \url{https://github.com/AMReX-Codes/amrex}, hash {\tt 3dda62} and this code can be obtained at
\url{https://github.com/AMReX-Microelectronics/FerroX}, hash {\tt 002bdd}.

\section*{Acknowledgements}
This work was supported by the US Department of Energy, Office of Science, Office of Basic Energy Sciences, the Microelectronics Co-Design Research Program, under contract no. DE-AC02-05-CH11231 (Codesign of Ultra-Low-Voltage Beyond CMOS Microelectronics) for the development of design tools for low-power microelectronics.
This research used resources of the National Energy Research Scientific Computing Center, a DOE Office
of Science User Facility supported by the Office of Science of the U.S. Department of Energy under Contract No. DE-AC02-05CH11231. This research leveraged the open source AMReX code, https://github.com/AMReX-Codes/amrex. We acknowledge all AMReX contributors.
The authors thank Lane Martin, Thomas Lee, Raul. A. Flores, Sinéad Griffin, and Ramamoorthy Ramesh for valuable discussions.
\section*{Author contributions}
P.K. and Z.Y. formulated the model, P.K., A.N., and Z.Y. developed the code, P.K. performed the simulations, P.K., Z.Y, and A.N. performed data analysis, G.P. and S.S. provided physical insights of the simulated results.  All authors discussed the results and prepared the manuscript.
\section*{Competing interests}
The authors declare no competing interests.

\noindent\textbf{References}
\bibliographystyle{naturemag}
\bibliography{Manuscript}

\end{document}